**Title:** Negative refraction of ultra-squeezed in-plane hyperbolic designer polaritons

*Qiaolu Chen, Yihao Yang,\* Li Zhang, Jialin Chen, Xiao Lin, Rujiang Li, Zuojia Wang, Baile Zhang,\* Hongsheng Chen\**

Qiaolu Chen, Li Zhang, Jialin Chen, Prof. Zuojia Wang, Prof. Hongsheng Chen
Interdisciplinary Center for Quantum Information, State Key Laboratory of Modern Optical Instrumentation, College of Information Science and Electronic Engineering, Zhejiang University, Hangzhou 310027, China.
E-mail: hansomchen@zju.edu.cn

Qiaolu Chen, Li Zhang, Jialin Chen, Prof. Zuojia Wang, Prof. Hongsheng Chen
ZJU-Hangzhou Global Science and Technology Innovation Center, Key Lab. of Advanced Micro/Nano Electronic Devices & Smart Systems of Zhejiang, Zhejiang University, Hangzhou 310027, China.
E-mail: hansomchen@zju.edu.cn

Dr. Yihao Yang, Dr. Xiao Lin, Dr. Rujiang Li, Prof. Baile Zhang
Division of Physics and Applied Physics, School of Physical and Mathematical Sciences, Nanyang Technological University, Singapore, Singapore.
E-mail: yang.yihao@ntu.edu.sg, blzhang@ntu.edu.sg

Dr. Yihao Yang, Dr. Rujiang Li, Prof. Baile Zhang
Centre for Disruptive Photonic Technologies, The Photonics Institute, Nanyang Technological University, Singapore, Singapore.
E-mail: yang.yihao@ntu.edu.sg, blzhang@ntu.edu.sg

**Keywords:** hyperbolic metasurfaces, negative refraction, hyperbolic polaritons

**Abstract:**

The in-plane negative refraction of high-momentum (i.e., high-$k$) photonic modes could enable many applications such as imaging and hyperlensing in a planar platform at deep-subwavelength scales. However, its practical implementation in experiments remains elusive so far. Here we propose a class of hyperbolic metasurfaces, which is characterized by an anisotropic magnetic sheet conductivity and can support the in-plane ultrahigh-$k$ magnetic designer polaritons. Based on such metasurfaces, we report the first experimental observation of the all-angle negative refraction of designer polaritons at extremely deep-subwavelength scales. Moreover, we directly visualize the designer polaritons with hyperbolic dispersions. Importantly, for these hyperbolic polaritons, we find that their squeezing factor is ultra-large and, to be specific, it can be up to 129 in the experiments, a record-breaking value exceeding those in naturally hyperbolic materials. The present scheme for the achievement of negative refraction is also applicable to other natural materials and may enable intriguing applications in nanophotonics. Besides, the proposed metasurfaces are readily tailorable in space and frequency, which could serve as a versatile platform to explore the extremely high confinement and unusual propagation of hyperbolic polaritons.

**Introduction:**

Three-dimensional (3D) hyperbolic media are characterized by the anisotropic permittivity or permeability tensor with a principal component being opposite in sign to the other two principal components.[1]-[4] One of the most important properties of hyperbolic media is that they support photonic modes with high momentum (one order larger than photons' momentum in free space (denoted as $k_0$)), allowing for confining photons at deep-subwavelength scales or even the extreme nanoscales. The high-momentum or high-$k$ modes play a key role in many applications of hyperbolic media,[1],[3] such as low-threshold Cherenkov radiation,[5]-[6] enhanced spontaneous emission,[7]-[8] super-Planckian thermal emission,[9]-[10] ultra-sensitive sensors,[11] subwavelength imaging,[12] and enhanced photon extraction.[13] Recent researches have revealed that such high-*k* modes also exist in ultrathin hyperbolic media, or two-dimensional (2D) hyperbolic media, such as hyperbolic metasurfaces (e.g., nanostructured van der Waals (vdW) materials,[14] designed metal-based metasurfaces),[15]-[21] and slabs of naturally hyperbolic materials[22]-[23] (e.g., the uniaxial boron nitride (BN),[24]-[27] the biaxial vdW crystal α-MoO$_3$).[28]-[29]

Realization of in-plane negative refraction of high-*k* modes is of great significance and interest, due to its unique applications such as sub-diffraction imaging/focusing, hyperlensing, and waveguiding in a planar platform at deep-subwavelength scales.[15],[30]-[33] Such a goal has become a possibility, thanks to the advent of artificial hyperbolic metasurfaces and naturally hyperbolic materials. Though the negative refraction of hyperbolic surface plasmon polaritons (SPPs) was previously experimentally demonstrated at an interface between a hyperbolic metasurface and a flat silver film, both the SPPs on the silver film and hyperbolic metasurface are relatively weakly confined (their wavevectors are usually less than 3$k_0$).[17] This approach cannot directly apply to the high-*k* modes, as the high-*k* surface-wave modes are usually strongly reflected or scattered when coupling to low-*k* modes (such as SPPs on the silver film), owing to the large in-plane and out-of-plane momentum mismatching.[34] Therefore, to achieve in-plane negative refraction of high-*k* modes, the dispersions in both regions need to be precisely engineered. One straightforward way to do so is to employ two materials/structures supporting high-*k* modes with opposite group velocities. For example, graphene supports high-*k* plasmon polaritons with positive group velocity and BN supports high-*k* phonon polaritons with negative group velocity in its first Reststrahlen band.[32] Moreover, by judiciously tuning the chemical potential of graphene and the thickness of BN, it is possible to flexibly flip the sign of the group velocity of the hybrid polaritons in graphene-BN heterostructures. As such, graphene-BN heterostructures can be a versatile platform to support the in-plane negative refraction of high-$k$ modes. However, the operational frequency bandwidth of negative refraction in graphene-BN heterostructures is within the first Reststrahlen band of BN and is thus very narrow. Such a strict requirement is then experimentally unfavorable and challenging. Recently, Jiang et al. proposed to realize the broadband all-angle negative refraction of high-$k$ hyperbolic plasmon polaritons at an interface constructed by two identical nanostructured graphene metasurfaces with different rotation angles.[33] However, both fabrication and experimental characterization of such graphene metasurfaces are quite challenging. In short, although there are some theoretical proposals of in-plane negative refraction of high-$k$ modes, their practical realization in experiments has not been reported so far.

To overcome the above challenges, we propose, design, and fabricate a class of hyperbolic metasurfaces characterized by an anisotropic magnetic sheet conductivity (denoted as $\sigma_m$),[35]-[36] which supports in-plane ultrahigh-$k$ magnetic designer polaritons. Based on such metasurfaces, all-angle negative refraction of ultrahigh-$k$ designer polaritons is observed in the experiments. Moreover, our microwave measurements directly show that the designer polaritons indeed exhibit in-plane hyperbolic dispersions and have ultra-squeezed wavelength down to 129 times smaller than the free-space photons' wavelength (denoted as $\lambda_0$). Such a record-breaking squeezing factor (defined as $\lambda_0/\lambda_p$, where $\lambda_p$ is the designer polaritons' wavelength) exceeds those in naturally in-plane hyperbolic materials demonstrated previously (usually less than 60 in the experiments).[28]-[29] Our work provides a distinctive way to achieve in-plane negative refraction of ultrahigh-$k$ modes, which could apply to optical in-plane hyperbolic materials, enabling many applications in nanophotonics. Additionally, our metasurfaces are readily tailorable in frequency and space, which form a highly variable platform for exploring the extremely high confinement and unusual propagation of in-plane hyperbolic polaritons.

**Results:**

The proposed anisotropic-$\sigma_m$ hyperbolic metasurface is composed of arrays of coiling copper wires patterned on dielectric substrates, as illustrated in Figure 1a. An enlarged view of a unit cell is shown in Figure 1b, where the grey region denotes the dielectric substrate with relative permittivity $\varepsilon_r = 3.5+0.001i$ and thickness $t = 2$ mm, and the orange region represents a 35-μm coiling copper wire with inner radius $r = 2$ mm, width $w = 0.2$ mm, gap $g = 0.2$ mm, and number of coil turns $N = 14$. The lattice constants are $a = 15.2$ mm ($x$ axis) and $b = 9$ mm ($y$ axis), respectively.

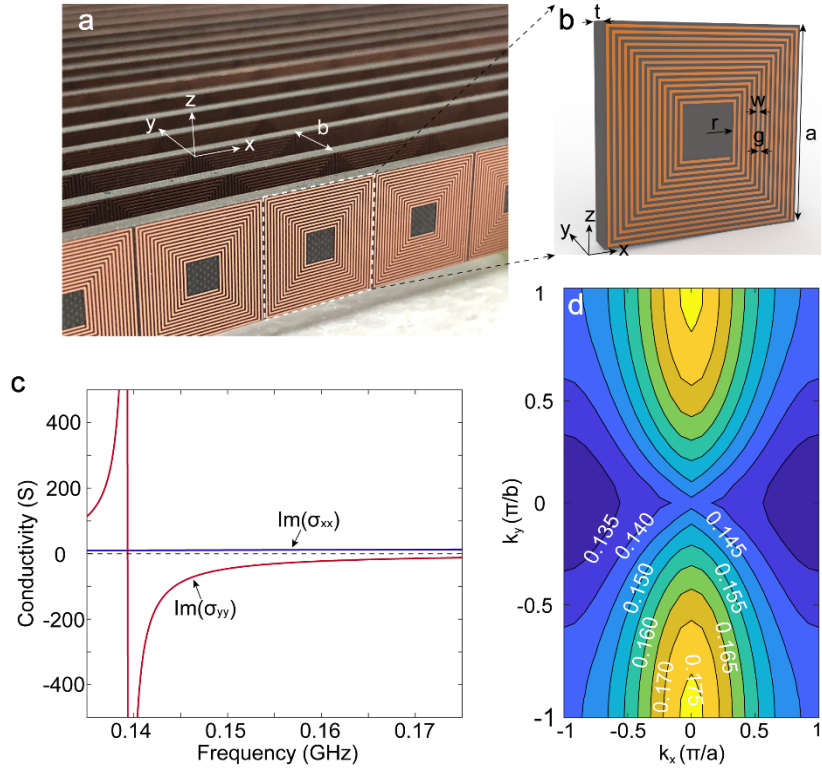

**Figure 1.** Anisotropic-$\sigma_m$ hyperbolic metasurface. a) Photograph of the fabricated sample. The sample is composed of arrays of coiling copper wires patterned on dielectric substrates. b) Details of a unit cell. Here, $a = 15.2$ mm, $b = 9$ mm, $r = 2$ mm, $t = 2$ mm, $w = 0.2$ mm, and $g = 0.2$ mm. The relative permittivity of substrate is 3.5+0.001i. The thickness of copper wires is 0.035 mm. Number of coil turns is 14. c) Numerically-calculated Im($\sigma_m$) of the metasurface. Blue (red) solid line represents values of Im($\sigma_{xx}$) (Im($\sigma_{yy}$)). Black dashed line denotes value of zero. d) Iso-frequency contours of the hyperbolic metasurface in the first Brillouin zone. The frequency values are presented in the unit of gigahertz.

We then numerically calculate the iso-frequency contours (IFCs) of the fundamental mode of the proposed hyperbolic metasurface in the first Brillouin zone (FBZ) by employing eigenvalue module of a commercial software Computer Simulation Technology (CST) Microwave Studio, as shown in Figure 1d. One can see that, the IFCs are open hyperbolas from 0.145 to 0.175 GHz. Note that the magnetic field is highly confined in both vertical and horizontal directions, indicating the surface-wave nature of the eigenmodes (see the Supporting Information). Besides, the operational wavelength is around $2\times10^3$ mm, which is much larger than the thickness of metasurface (i.e., 15.2 mm).

To understand the behaviors of the designer polaritons over the metasurface, we model the proposed hyperbolic metasurface as a 2D magnetic sheet conductivity layer surrounded by two vacuum half-spaces.[37] Here we consider the hyperbolic metasurface as a lossless medium, because ohmic losses of the substrate and copper are negligible in the microwave region. Then, by adopting a standard retrieval method,[35] we extract the magnetic sheet conductivity of the metasurface

$$\overline{\overline{\sigma_m}} = \begin{pmatrix} \sigma_{xx} & 0 \\ 0 & \sigma_{yy} \end{pmatrix}. \qquad (1)$$

The resulting $\sigma_m$ plot is shown in Figure 1c, where Im($\sigma_{xx}$) and Im($\sigma_{yy}$) are opposite in sign from 0.145 to 0.175 GHz, indicating a hyperbolic frequency band. Note that, TM modes here are barely confined on the metasurface, thus we mainly focus on TE modes in our case, and the corresponding dispersion relation of the designer polaritons is

$$\frac{1}{\eta_0^2}(\sigma_{xx}k_x^2 + \sigma_{yy}k_y^2)^2(k_x^2+k_y^2-k_0^2) + 4k_0^2(k_x^2+k_y^2)^2 = 0, \qquad (2)$$

where $k_{x,y}$ are the in-plane wavevectors, and $\eta_0$ is the characteristic impedance of vacuum.[37]-[38] Comparing the analytical and simulated dispersions, we find a good agreement between them (see the Supporting Information).

In the following, we carry out experiments to characterize the proposed hyperbolic metasurface. The experimental sample consists of 20 by 32 unit cells, taking up an area of 304 mm by 288 mm. To excite the designer polaritons efficiently, a port of the vector network analyzer (VNA) is directly connected to a coil unit cell at the edge of the metasurface. Another port of the VNA is connected to a detector that is a compact coil with a magnetic resonance around 0.159 GHz. The coil-like detector that oriented in the z direction, is fixed at a robotic arm of a movement platform and moves on the *xy* plane 3 mm above the metasurface. By scanning the sample, the complex magnetic patterns of $H_z$ field (including phase and amplitude) are recorded. The size of the scanning region is approximately 250 mm by 250 mm, with a resolution of 15.2 mm by 9 mm. See the details of the experimental set-up in the Supporting Information.

The measured field patterns are shown in Figure 2a. One can directly observe the significant feature of hyperbolic designer polaritons, that is, concave polariton wavefronts. For further proof of in-plane hyperbolic dispersions, we extract the IFCs in the momentum space by applying spatial Fourier transform to the corresponding complex field patterns (see Figure 2b). One can see that the measured IFCs are indeed hyperbola-like curves, thus experimentally corroborating the simulated IFCs (Figure 1d). Note that there is a bright spot at the center of the FBZ at each frequency, which attributes to the radiation noise.[21]

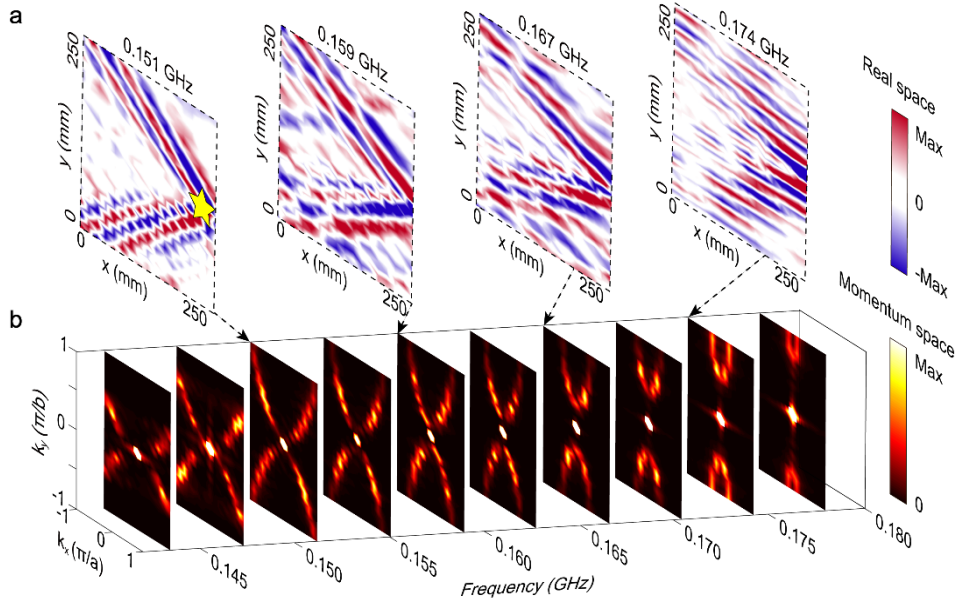

**Figure 2.** Measured magnetic field distributions and iso-frequency contours. a) Measured magnetic patterns of $H_z$ field in the real space on the $xy$ plane 3 mm over the hyperbolic metasurface at 0.151, 0.159, 0.167, and 0.174 GHz, respectively. The yellow star indicates the source position. The colorbar measures the real part of $H_z$. b) Iso-frequency contours in the momentum space obtained by applying spatial Fourier transform to the corresponding complex $H_z$ field. The colorbar measures the energy intensity.

Next, we retrieve the squeezing factors of the hyperbolic designer polaritons, which are determined by the largest polaritons' wavevector in the FBZ at each frequency. As shown in Figure 3a, the squeezing factors of the in-plane polaritons are remarkably large, with a maximum value of 129 in both experiments and simulations, revealing the ultrahigh-$k$ nature of the hyperbolic designer polaritons. Note that, compared with the squeezing factors in the previous in-plane hyperbolic metasurfaces or naturally in-plane hyperbolic materials such as vdW crystal $\alpha$-MoO$_3$, which are usually less than 60,[28]-[29] the maximum squeezing factor achieved in our experiments is a record-breaking high value. Figure 3b depicts the IFC directly obtained by applying spatial Fourier transform to the corresponding complex field pattern of $H_z$ at 0.150 GHz (Figure 3c) where the squeezing factor is maximal.

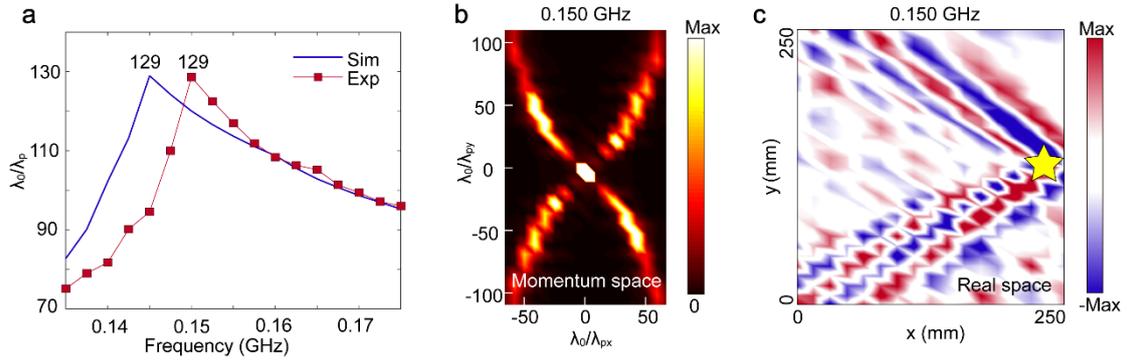

**Figure 3.** Achieving a record-breaking high squeezing factor in our hyperbolic metasurface. a) Retrieved squeezing factors of the designer polaritons from simulated and experimental results. A record-breaking high squeezing factor of 129 at 0.150 GHz is observed. b) Measured iso-frequency contour at 0.150 GHz. The colorbar measures the energy intensity. c) Measured $H_z$ field on the *xy* plane 3 mm over the hyperbolic metasurface at 0.150 GHz. The excitation is marked as a yellow star. The colorbar measures the real part of $H_z$.

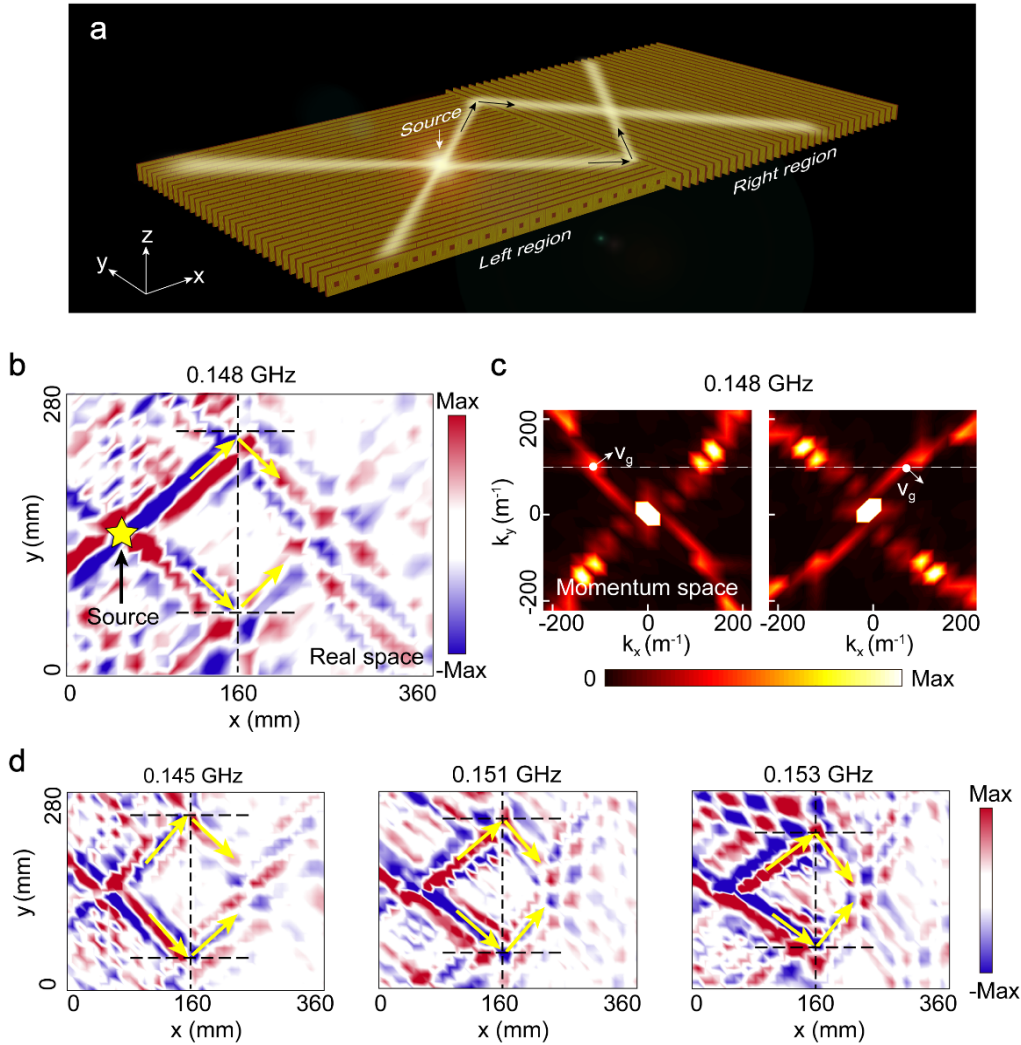

**Figure 4.** Experimental validation of all-angle in-plane negative refraction of ultrahigh-$k$ designer polaritons. a) Schematic view of in-plane negative refraction. The hyperbolic metasurface in the right region is rotated by 90°, comparing with the left one. The black arrows indicate the power flow. b) Measured $H_z$ field on the $xy$ plane 3 mm above the hyperbolic metasurface at 0.148 GHz. The vertical black dashed line denotes the interface. The horizontal black dashed lines are normals. The excitation is marked as a yellow star. The yellow arrows represent the directions of the power flow. The colorbar measures the real part of $H_z$. c) Iso-frequency contours of the hyperbolic metasurfaces in the left and right regions at 0.148 GHz, respectively. The white arrows indicate the directions of group velocities (power flow). The colorbar measures the energy intensity. d) Measured $H_z$ field at 0.145, 0.151, 0.153 GHz, respectively.

Finally, we realize the in-plane negative refraction of the designer polaritons, based on the present anisotropic-$\sigma_m$ hyperbolic metasurface. We first construct an interface consisting of two identical metasurfaces apart from a 90° rotation (see the schematic in Figure 4a). A point source is placed at the center of the left region and hyperbolic designer polaritons with concave wavefronts are launched. Remarkably, the designer polaritons negatively refract at the interface, with the incident and refracted beams at the same side of the normal, as shown in Figure 4b. This intriguing phenomenon can be explained by the measured IFCs of both regions, as illustrated in Figure 4c. One can see that the directions of $y$-component of group velocities (parallel to the interface) are opposite for incident and refracted polaritons, according to the conservation law where the tangential components of the wavevectors in left and right regions should be matched. As the group velocity determines the direction of power flow, negative refraction of designer polaritons occurs at the interface.[33] Note that, since the polaritons' momentum is much larger than $k_0$, negligible energy leaks into the air at the interface. However, for low-$k$ surface-wave modes, to prevent the leakage, the out-of-plane wavevectors in two regions should be carefully matched.[34]

Additionally, the negative refraction also occurs at other frequencies, e.g., 0.145, 0.151, and 0.153 GHz, as shown in Figure 4d, which indicate the relatively broad operational bandwidth. We would like to mention that for any designer polaritons launched in the left/right region, the negative refraction always happens at the interface regardless of the incidence angle. Therefore, we denote such a phenomenon as all-angle in-plane negative refraction.[32]-[33],[39]-[41]

**Discussion:**

We have identified a class of anisotropic-$\sigma_m$ hyperbolic metasurfaces, which supports ultrahigh-$k$ in-plane magnetic designer polaritons featured with hyperbolic dispersions. Based on the proposed metasurface, all-angle in-plane negative refraction of the ultra-squeezed designer polaritons is observed experimentally. Remarkably, a record-breaking squeezing factor of 129 is achieved in the experiments, which exceeds those in the naturally or artificially in-plane hyperbolic materials demonstrated previously. The all-angle in-plane negative refraction of ultrahigh-$k$ hyperbolic designer polaritons proposed here could also be applicable to the optical hyperbolic materials,[14],[28]-[29] enabling novel applications in nanophotonics. Besides, the present metasurfaces with highly squeezing factors and tunability, could serve as an excellent platform to explore the applications and physics of in-plane hyperbolic polaritons. Finally, by scaling down the unit cell, our design could apply to higher frequencies (see a design in the Supporting Information), such as terahertz and far-infrared frequencies, and may find many applications in flatland optics,[31] e.g., waveguiding, imaging, giant thermal emission, and enhanced spontaneous emission.

**Supporting Information**
Supporting Information is available from the Wiley Online Library or from the author.

**Acknowledgements**
Q.C. and Y.Y. contributed equally to this work. The work at Zhejiang University was sponsored by the National Natural Science Foundation of China (NNSFC) under Grants No. 61625502, No.11961141010, and No. 61975176, the Top-Notch Young Talents Program of China, the Fundamental Research Funds for the Central Universities. This work at Nanyang Technological University was sponsored by Singapore Ministry of Education under Grant Nos. MOE2018-T2-1-022 (S), MOE2015-T2-1-070, MOE2016-T3-1-006, and Tier 1 RG174/16 (S).

Received: ((will be filled in by the editorial staff))
Revised: ((will be filled in by the editorial staff))
Published online: ((will be filled in by the editorial staff))